\documentclass[twocolumn,aps,amsmath,nofootinbib,superscriptaddress]{revtex4}
\usepackage{psfig}

\begin{document}

\title{The Casimir Effect in the Presence of Compactified Universal Extra Dimensions}

\author{Katja Poppenhaeger}
\email{katjap@th.physik.uni-frankfurt.de}
\affiliation{  
 Institut f\"ur Theoretische Physik\\ 
J. W. Goethe-Universit\"at\\
Robert-Mayer-Str. 8-10\\ 
60054 Frankfurt am Main, Germany}
\author{Sabine Hossenfelder}
\affiliation{  
 Institut f\"ur Theoretische Physik\\ 
J. W. Goethe-Universit\"at\\
Robert-Mayer-Str. 8-10\\ 
60054 Frankfurt am Main, Germany}
\author{Stefan Hofmann}
\affiliation{  
Department of Physics\\
Stockholm University\\
SE-106 91 Stockholm, Sweden}
\author{Marcus Bleicher}
\affiliation{  
 Institut f\"ur Theoretische Physik\\ 
J. W. Goethe-Universit\"at\\
Robert-Mayer-Str. 8-10\\ 
60054 Frankfurt am Main, Germany}

\begin{abstract}
The Casimir force in a system consisting of two parallel conducting plates in the presence of
compactified universal extra dimensions (UXD) is analyzed. The Casimir force with UXDs differs from the force obtained without extra dimensions. A new power law for the Casimir 
force is derived. By comparison to experimental
data the size $R$ of the universal extra dimensions can be restricted to  $R \leq 10 \,{\rm
nm}$ for one extra dimension.
\vspace{1cm}

\end{abstract}

\maketitle

The idea that our world has more than three spatial dimensions has been discussed for more than
80 years. Already Kaluza and Klein \cite{Kaluza:1984ws,Klein:tv} postulated an additional compactified 
dimension to unify gravity and classical electrodynamics. Today, there is a large variety of promising 
models and theories which suggest the existence of more than three spatial dimensions. 
Most notably, string theory \cite{Green:sp,Horava:1995qa} suggests the existence of  seven additional 
spatial dimensions. In string theory (and also in older approaches) one expects that the 
compactification scale of the extra dimensions is of order $M_{{\rm Planck}}\sim 10^{19}$~GeV.
Thus, observable effects are shifted into an energy domain out of reach of todays and near future 
experimental possibilities.

However, recently models with compactification radii up to the mm-scale and with a lowered Planck-Mass 
(TeV region) have been introduced \cite{Antoniadis:1990ew,Antoniadis:1998ig,Arkani-Hamed:1998nn,Kakushadze:1998wp,Dienes:1998qh,Randall:1999vf,Randall:1999ee,Appelquist:2000nn}. 
In the ADD \cite{Arkani-Hamed:1998nn} and RS \cite{Randall:1999vf} type of models, the hierarchy-problem is solved
or reformulated, resp. in a geometric language. Here, the existence of $d$ compactified large extra dimensions 
({\sc LXD}s, because the size of the additional dimensions can be as large as $100-1000 \mu$m) in which 
only the gravitons can propagate is assumed. The standard-model 
particles however, are bound to our 4-dimensional submanifold, often called our 3-brane.

This obvious asymmetry between standard model fields and gravity has given rise to the introduction
of universal extra dimensions ({\sc UXD}s) in which all particle species are allowed to propagate 
\cite{Antoniadis:1990ew,Appelquist:2000nn}. In this model the present limit on the size of the extra dimensions
is $R\le (300 \, {\rm GeV})^{-1} \approx 10^{-9}\, {\rm nm}$ due to the non-observation of Kaluza-Klein 
excitations at Tevatron \cite{Appelquist:2000nn,Macesanu:2002ew,Rizzo:2001sd,Wells:2002gq}.

Especially the Casimir effect has received great attention and has been extensively studied in a 
wide variety of topics in those and related scenarios:
\begin{itemize}
\item The question how vacuum fluctuations affect the stability of extra dimensions has been
explored in \cite{Ponton:2001hq,Hofmann:2000cj,Huang:2000qc,Graham:2002xq,Saharian:2002bw,Elizalde:2002dd}. 
Especially the detailed studies  in the Randall-Sundrum model have shown the major contribution of
the Casimir effect to stabilise the radion \cite{Garriga:2002vf,Pujolas:2001um,Flachi:2001pq,Goldberger:2000dv}. 
\item Cosmological aspects like the cosmological constant as a manifestation of the Casimir energy 
or effects of Casimir energy during the  primordial cosmic inflation have been analyzed 
\cite{Peloso:2003nv,Elizalde:2000jv,Pietroni:2002ey,Setare:2003ds,Melissinos:2001fm,Gardner:2001fz,Milton:2001np,Carugno:1995wn,Naylor:2002xk}. 
\item The Casimir effect in the context of string theory has been investigated 
in \cite{Fabinger:2000jd,Gies:2003cv,Brevik:2000fs,Hadasz:1999tr}.
\end{itemize}

The commonly known and experimentally accessible Casimir effect \cite{Casimir:dh} has 
recently gained intense attention. Experimentally \cite{Lamoreaux:1996wh} the precision of the
measurements has been greatly enhanced, while on the theoretical side major progress 
has been reported (for a review, the reader is referred to \cite{Bordag:2001qi}). In
fact, the Casimir effect has been suggested as an
experimentally powerful tool for the investigation of new physics beyond the
standard model \cite{Krause:1999ry}. 

In this letter, we scrutinize the Casimir force between two 
parallel plates to probe the possible existence and size of additional Universal 
Extra Dimensions (UXDs) \cite{Appelquist:2000nn}. 
In general and independently from the considered field, zero-point fluctuations of any quantum field give rise to
observable Casimir forces if boundaries are present \cite{Casimir:dh}. In this study, 
the zeta function method as suggested in 
\cite{Beneventano:1995fh,Svaiter:je,Gosdzinsky:1998qa,Leseduarte:1996ah,Cognola:1999zx,Li:qf,Bordag:1995zc} is applied to 
renormalize the Casimir energy.
For simplicity, we limit ourselves to the exploration of an UXD model with only one extra dimension 
\cite{Appelquist:2000nn} which is compactified on a $S^1/Z_2$ orbifold. 
This restricts the possible vacuum fluctuations of the electromagnetic field to have a 
wave  vector along the extra dimension of the form $k_n=n/R$,
with $k_n$ being the wave vectors in direction of the universal extra dimension, $n$ an integer and
$R$ the radius of the extra dimension.
Because we study the behaviour of the Casimir energy in a system with parallel conducting
plates in the presence of UXDs, we also need the boundary condition
$k_N=\pi N/a$,
where $k_N$ is the wave vector in the directions restricted by the plates, $N$ an integer and $a$
the distance of the plates.

In the case of one extra dimension we find the frequency of the vacuum fluctuations
to be
\begin{eqnarray}
\omega_{nN} = c\sqrt{k_\bot ^2 + \frac{n^2}{R^2}+\left(\frac{\pi
N}{a}\right)^2}\quad ,
\end{eqnarray}

with $k_\bot=\sqrt{\left(k_1^2+k_2^2\right)}$. $k_1$ and $k_2$ are the wave vectors in direction of
the unbound space coordinates.  Therefore the Casimir energy per unit plate area reads
\begin{eqnarray}
\varepsilon_{{\rm nr}} =
2\cdot\frac{\hbar }{2}\int\frac{{\rm d}^2 k_\bot}{(2\pi)^2}\left(
{\sum_{n,N=0}^{+\infty'}}p \cdot\omega_{nN}-{\sum_{n=0}^{+\infty'}}\omega_{n0} \right)
\end{eqnarray}

Here, the prime indicates that the term with $n=N=0$ has to be skipped. The factor $2$
arises from the volume of the orbifold, and the factor $p$ from the possible polarizations of the photon ($p=3$ for one UXD). The modes polarized in the direction of our brane, $n=0$, cause the additional term which has to be subtracted \cite{Ambjorn:1981xw}.

Using the Schwinger representation \cite{Schwinger:nm} of the square root and the
Gauss integral in $2$ dimensions one obtains
\begin{eqnarray}
\varepsilon_{{\rm nr}} &=&
\frac{\hbar c}{4 \pi \Gamma\left(-\frac{1}{2}\right)}
  \int\limits_0^\infty\frac{{\rm d}x}{x} x^{-3/2}\nonumber\\
&\times&\left[p{\sum_{n,N=0}^{+\infty'}} \exp{\left(-\frac{n^2}{R^2}-\left(\frac{\pi
N}{a}\right)^2\right)x}\right.\nonumber\\
&-&\left.{\sum_{n=0}^{+\infty'}} \exp{\left(-\frac{n^2}{R^2}x\right)}\right]\quad .
\end{eqnarray}
\\
With the Gamma-function \cite{abramowitz} this yields
\begin{eqnarray}
\varepsilon_{{\rm nr}} &=&
\frac{\hbar c}{4\pi\Gamma\left(-\frac{1}{2}\right)}
\Gamma\left(-\frac{3}{2}\right)\nonumber\\
&\times&\left[
{p\sum_{n,N=0}^{+\infty'}}
\left(\frac{n^2}{R^2}+\left(\frac{\pi
N}{a}\right)^2\right)^{3/2}-{\sum_{n=0}^{+\infty'}}
\left(\frac{n^2}{R^2}\right)^{3/2}\right]\quad .
\end{eqnarray}

To use the Epstein zeta function \cite{Ambjorn:1981xw} renormalization, the sums are re-written 
with indices running from $-\infty$ to $+\infty$. For the double sum, one has to take 
care of the modes where one index equals zero. 
The resulting energy density reads in terms of the Epstein zeta-function $Z$:
\begin{eqnarray}
\varepsilon_{{\rm nr}} &=&
\frac{\hbar c}{4\pi\Gamma\left(-\frac{1}{2}\right)}
\Gamma\left(-\frac{3}{2}\right)
\left[\frac{p}{4}Z_2\left(\frac{1}{R}, \frac{\pi}{a},-3\right)\right.\nonumber\\
&+&
\left.\frac{p-2}{4}Z_1\left(\frac{1}{R}, -3\right)+\frac{p}{4}Z_1\left(\frac{\pi}{a}, -3\right)\right]\quad .
\end{eqnarray}

\begin{figure}
\psfig{file=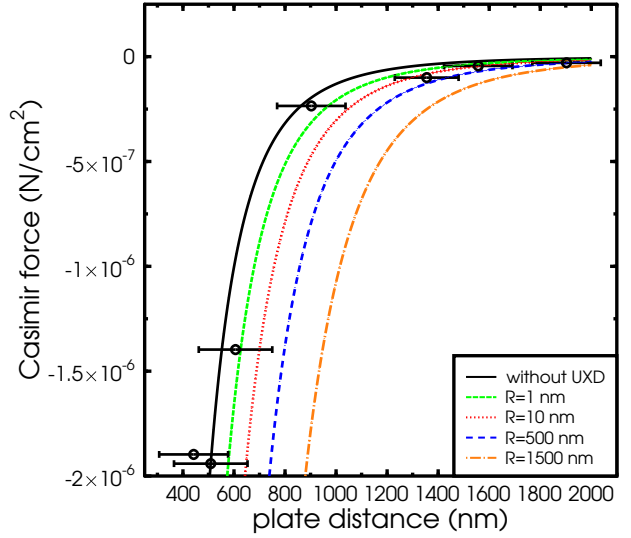,width=9cm}
\caption{\label{casiforce} Casimir
forces with (dashed and dotted lines) and without (full line) UXD as a function of the
plate separation. The symbols denote data taken from \cite{Sparnaay:1958wg}.}
\end{figure}

\vspace{0.5cm}

The second and the third term are due to the modes with one index equal to zero.
According to \cite{Ambjorn:1981xw}, the reflexion relation of the Epstein zeta 
function yields the renormalized energy density:

\begin{eqnarray}
\varepsilon_{\rm ren} &=&
-\frac{ \hbar c}{16\pi^5}
\left[\frac{3}{8\pi}p\cdot R a
Z_2\left(R,\frac{a}{\pi},5\right)\right.\nonumber\\
&+&
\left.\frac{p-2}{R^3}\zeta(4)+\frac{p}{a^3}\pi^3\zeta(4)\right]\quad .
\end{eqnarray}

Note that this quantity is regularized with respect to $3+1$-dimensional Minkowski space. To
obtain the total energy in the space between the plates, one has to multiply by the surface $A$ of the plates:
$E(R,a)=\varepsilon_{{\rm ren}}\cdot A$.

In an analogous calculation we find the renormalized vacuum energy in the
volume between the plates, but plates absent to be
\begin{eqnarray}
E_{{\rm UXD}}(R)&=&-\frac{\hbar
c}{16\pi^5}\frac{3}{4\pi}p\cdot a\frac{1}{R^4}\zeta(5)\cdot A\quad .
\end{eqnarray}

The quantity of interest for the Casimir effect in the present setting is the energy difference
between $3+1$-dimensional space with plates and $3+1$-dimensional space without plates. 
Thus, in the setting with one compactified universal extra dimension the Casimir energy is
\begin{eqnarray}
E_{{\rm cas}}(R,a)=E_{{\rm UXD}}^{\rm Plates}(R,a)- E_{{\rm UXD}}(R,a) \quad.
\end{eqnarray}

The Casimir force is now given by the derivative of the Casimir energy $E_{{\rm cas}}$
with respect to the plate distance $a$:
\begin{eqnarray}
F_{{\rm cas}}&=&-\frac{\partial E_{{\rm cas}}}{\partial a}\nonumber\\
&=& \frac{\hbar c}{16 \pi^5}pA \left[\frac{3}{8\pi}R \cdot Z_2\left(R,\frac{a}{\pi},5\right) \right.
\nonumber\\
&-&\frac{15}{8\pi^3}a^2R \sum_{n,N=-\infty}^{+\infty'}
N^2\left[n^2R^2+\frac{a^2}{\pi^2}N^2\right]^{-\frac{7}{2}}\nonumber\\
&-&\left. 3\pi^3\zeta(4)\frac{1}{a^4}-\frac{3}{4\pi}\zeta(5)\frac{1}{R^4} \right]
\end{eqnarray}

Now we compare the Casimir force in this modified space-time to data and the normal Casimir force $F$ between
parallel plates without extra dimensions, given by
\begin{eqnarray}
F =-\frac{\hbar c\pi ^2}{240}\cdot \frac{A}{a^4}\quad .
\end{eqnarray}

It should be noted that the measurement of Casimir forces between parallel plates is experimentally 
difficult because exact parallelity cannot be 
obtained easily\footnote{Today, Casimir forces are mostly measured in settings
with a plate and a sphere because one has not to deal with the problem of parallelity. The first
high-precision experiment with plate-sphere setting is given in \cite{Lamoreaux:1996wh}. 
For newer results the reader is referred to \cite{Mohideen:1998iz,Roy:1999dx,Bressi:2002fr}.}.
In spite of this problem, one experiment with relatively high accuracy was done by
Sparnaay \cite{Sparnaay:1958wg} with chromium plates\footnote{Unfortunately, Sparnaay's measurement 
with other metals showed partly repulsive
instead of attractive forces, so the present experimental results should be handled with care.}.

Figure \ref{casiforce} depicts the dependence of the Casimir force between two parallel plates
on the radius of the extra dimension (dashed and dotted lines) and the distance of the plates. 
One clearly observes that the data can be reproduced either by a calculation without {\sc UXD}s (full line)
or by a calculation with one {\sc UXD} of small size.
In the present setting with one universal extra dimension, good agreement with the
data can only be obtained if the radius of the UXD is smaller than $\sim 10{\rm
nm}$. Similar results are obtained for more than one Universal Extra Dimension.  

In conclusion, the Casimir force between two conducting plates in the presence of {\sc UXD}s is studied.
For {\sc UXD} sizes previously discussed in the literature \cite{Appelquist:2000nn,Macesanu:2002ew,Rizzo:2001sd},
$R\approx (300 \, {\rm GeV})^{-1} \approx 10^{-9}\, {\rm nm}$, the Casimir force is in line
with the measured data. The present study of the Casimir force with one UXD yields an upper limit of $R\leq 10$ nm on the extension of UXDs.\\[.3cm]

\noindent
{\bf Acknowledgement}\\
The authors thank T. Bringmann, E. Ponton and H. St\"ocker for fruitful discussions. This work has been
supported by the German BMBF, GSI and DFG. S.~Hofmann acknowledges financial support from the
Wenner-Gren Foundation. S.~Hossenfelder wants to thank the Land Hessen for financial support.
K. Poppenhaeger acknowledges support from the
Studienstiftung des deutschen Volkes (German national merit foundation).

\end{document}